\documentclass[fleqn,usenatbib]{mnras}

\usepackage{newtxtext,newtxmath}
\usepackage[T1]{fontenc}
\usepackage{graphicx}
\usepackage{amsmath}
\usepackage{booktabs}
\usepackage{microtype}
\usepackage{url}
\usepackage{balance}

\newcommand{\Msun}{\mathrm{M}_{\odot}}
\newcommand{\Mearth}{\mathrm{M}_{\oplus}}
\newcommand{\Rearth}{\mathrm{R}_{\oplus}}
\newcommand{\kms}{\mathrm{km\,s^{-1}}}
\newcommand{\vimp}{v_{\rm imp}}
\newcommand{\ventry}{v_{\rm entry}}
\newcommand{\alttext}[1]{\par\smallskip{\footnotesize\noindent\textbf{Alt text:} #1\par}}

\title[Planetary eccentricity and comet delivery]{Planetary eccentricity enhances inward comet transport but suppresses gentle impacts}

\author[Q. Jiang]{
Qunfeng Jiang$^{1}$\thanks{E-mail: qfjiang01@gmail.com; ORCID: 0009-0008-0600-3743}\\
$^{1}$Independent Researcher, Shanghai, China
}

\date{Accepted XXX. Received YYY; in original form ZZZ}
\pubyear{2026}

\begin{document}
\label{firstpage}
\pagerange{\pageref{firstpage}--\pageref{lastpage}}
\maketitle

\begin{abstract}
Comets can supply volatile and prebiotic material to rocky planets, but survival of fragile molecules depends on atmospheric entry and impact. Compact planet chains can transfer comets inward at low velocity. We test whether modest planetary eccentricity preserves this route. We follow 64,000 comets from the region between the two outermost planets of an eight-planet chain to an Earth-mass planet on a temperate orbit around a $0.1\Msun$ star. A second suite of simulations isolates the last transfer to the innermost planet across initial planet eccentricities $e_{\rm p,0}=0$--0.05 and separations of 10 and 30 mutual Hill radii; a $\Delta=12$ comparison tests the influence of the near-3:2 period ratio at $\Delta=10$. At $e_{\rm p,0}=0.05$, the fraction of comets intercepted by the outermost planet falls by 11 per cent and 6 per cent more reach the inner region. The innermost planet's impact probability changes by less than 3 per cent, while the fraction of impacts below $15\,\kms$ falls from 25 to 18 per cent. The fraction of all injected comets producing impacts below $15\,\kms$ consequently decreases by 30 per cent. Measurements of the comet--planet speed before the final close encounter show that eccentricity preferentially removes the slowest approaches. Applying an impact-chemistry model gives a 31 per cent lower HCN survival-equivalent yield. Modest planetary eccentricity can therefore increase inward transport while reducing the prebiotic material delivered under chemically favourable impact conditions.
\end{abstract}


\section{Introduction}
\label{sec:introduction}

Material formed beyond the snow line can supplement the volatile and organic inventories of rocky planets after their assembly. Dynamical models of the Solar system place water-rich planetesimals in several outer reservoirs and show that planetary growth and scattering can transport a fraction of them into the terrestrial region \citep{Morbidelli2000,RaymondIzidoro2017}. Cometary nuclei also retain a chemically diverse inventory, including hydrogen cyanide (HCN), simple organics and larger carbon- and nitrogen-bearing species \citep{MummaCharnley2011,Altwegg2017}. These materials connect planet formation to surface environments in which prebiotic chemistry may proceed.

The chemical value of an impact depends on its physical conditions. HCN is a feedstock for cyanosulfidic reaction networks that produce precursors of nucleotides, amino acids and lipids \citep{Patel2015}. Wet--dry cycling can concentrate delivered material and promote subsequent polymerisation \citep{Pearce2017}. Impact heating, however, destroys HCN increasingly efficiently as velocity and incidence angle rise \citep{ToddOberg2020,McDonald2025}. Atmospheric passage introduces an additional size-dependent filter before an impactor reaches the surface \citep{Anslow2025}. An impact count therefore does not determine the amount of chemically useful material delivered to a planet.

A chain of planets can transport small bodies from a cold outer reservoir to the inner system. During scattering by one planet on a circular orbit, approximate conservation of the Tisserand parameter restricts the particle's accessible orbits \citep{MurrayDermott1999,BonsorWyatt2012}. Successive encounters with neighbouring planets reset that constraint and allow the pericentre to move inward in stages. Low-mass planets are less efficient ejectors, and close spacing increases the overlap between adjacent scattering zones \citep{Wyatt2017}. Numerical calculations consequently find strong links between planet mass and spacing, inward flux, ejection and planetary accretion \citep{Marino2018}. Architectures that maximise inner-system arrival can have lower accretion efficiencies because fast passage through a tightly packed chain reduces the time available for collision.

Impact velocity adds another dimension to this transport problem. A compact circular chain can pass a comet inward while preserving similar orbital velocity vectors through successive planet-to-planet transfers, producing a low-speed tail at the innermost planet \citep{Anslow2023}. This channel is especially consequential around low-mass stars. Their temperate orbits lie close to the star, so the local Keplerian speed is high; a small loss of velocity coherence can move an impact out of the narrow regime favourable to molecular survival. Compact terrestrial systems around very low-mass stars demonstrate that the required planetary multiplicity exists in nature \citep{Gillon2017}, and the regular sizes and spacings of many Kepler systems motivate an equal-mass, mutual-Hill-spaced model as a controlled dynamical laboratory \citep{Weiss2018}.

Circular orbits provide the most coherent possible planet-to-planet transfer. Observed compact systems are dynamically cold, yet eccentricities of a few per cent are common and often measurably non-zero \citep{HaddenLithwick2017,VanEylen2019}. Stability studies also show that eccentricity vectors and apsidal geometry matter in closely spaced systems \citep{GratiaLissauer2020}. At a representative temperate orbit around a $0.1\Msun$ star, the Keplerian speed is approximately $50\,\kms$. An eccentricity of 0.03 then introduces orbital-velocity variations of order $e v_{\rm K}\simeq1.5\,\kms$, comparable to the width of the slow impact tail. This additional velocity scale can alter comet transport and the delivery of impact-sensitive prebiotic material, motivating a direct test of gentle delivery in slightly eccentric compact systems.

This study follows comet transport through a complete compact chain and resolves the encounter that sets the final impact speed. The end-to-end calculation measures the probability of a gentle impact per comet supplied between the two outermost planets. Controlled final-transfer calculations determine how the velocity response depends on eccentricity and planet spacing. These calculations link modest eccentricity to both delivery efficiency and chemical value.

The paper is organised as follows. \hyperref[sec:methods]{Section~\ref*{sec:methods}} presents the planetary model, numerical integrations and diagnostics. \hyperref[sec:results]{Section~\ref*{sec:results}} describes the transport, impact yields and encounter velocities. \hyperref[sec:discussion]{Section~\ref*{sec:discussion}} considers the dynamical and astrobiological implications, and \hyperref[sec:conclusions]{Section~\ref*{sec:conclusions}} summarises the main findings.

\section{Methods}
\label{sec:methods}

\subsection{Planetary system}
\label{sec:architecture}

The fiducial system contains a $0.1\Msun$, $0.12\mathrm{R}_{\odot}$ star and eight coplanar planets, each with mass $1\Mearth$ and radius $1\Rearth$. The stellar mass focuses the calculation on the high orbital speeds characteristic of temperate planets around late M dwarfs. The innermost planet, P1, is placed at 0.0355 au, where its circular Keplerian speed is $49.99\,\kms$. This orbital scale lies in the temperate region of standard low-mass-star habitable-zone calculations \citep{Kopparapu2013}. The chain extends beyond 0.20 au, a representative snow-line scale for a low-mass protoplanetary disc \citep{KennedyKenyon2008}.

Adjacent planets are separated by $\Delta$ mutual Hill radii,
\begin{equation}
 a_{i+1}-a_i=\Delta R_{\rm H,m}, \qquad
 R_{\rm H,m}=\left(\frac{2m_{\rm p}}{3M_\star}\right)^{1/3}
 \frac{a_i+a_{i+1}}{2},
 \label{eq:mutualhill}
\end{equation}
where $a_i$ and $a_{i+1}$ are adjacent semimajor axes, $m_{\rm p}$ is the planet mass and $M_\star$ is the stellar mass. The fiducial value $\Delta=10$ gives semimajor axes from 0.0355 to 0.24037 au and an adjacent period ratio of 1.507.

The end-to-end experiment compares $e_{\rm p,0}=0$ and 0.05. In the eccentric systems all planets begin with the same eccentricity magnitude and independent, uniformly distributed longitudes of pericentre. Sixteen sets of planetary longitudes are generated. Each eccentric architecture is paired with a circular architecture that uses the same mean longitudes and the same comet initial conditions. Mutual planetary perturbations are included throughout the integration.

\subsection{End-to-end delivery experiment}
\label{sec:endtoend_methods}

\begin{figure*}
\includegraphics[width=\textwidth]{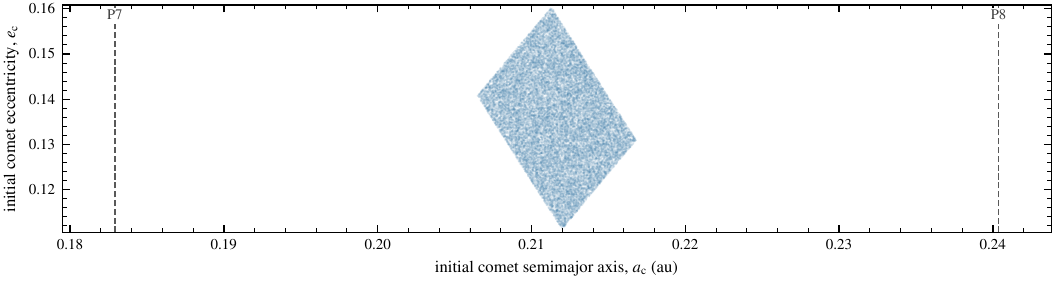}
\caption{Initial $a_{\rm c}$--$e_{\rm c}$ distribution of the 32,000 comets used in each end-to-end calculation. Dashed lines mark the semimajor axes of P7 and P8. The circular and eccentric planet calculations use identical comet initial states.}
\label{fig:initial_conditions}
\alttext{The comet initial states form a compact diamond-shaped distribution between P7 and P8, spanning eccentricities from 0.112 to 0.160.}
\end{figure*}

The first experiment measures the complete pathway from the outer edge of the planet chain to P1. Comets begin on transfer orbits between P7 and P8. Their pericentre $q$ and apocentre $Q$ are drawn independently from uniform distributions over
\begin{equation}
 q=(0.97\mbox{--}1.03)a_7, \qquad
 Q=(0.98\mbox{--}1.02)a_8.
 \label{eq:comet_window}
\end{equation}
The corresponding semimajor axis and eccentricity are
\begin{equation}
 a_{\rm c}=\frac{q+Q}{2}, \qquad
 e_{\rm c}=\frac{Q-q}{Q+q}.
 \label{eq:comet_elements}
\end{equation}
The argument of pericentre is uniform on $[0,2\pi)$, and the initial true anomaly lies within 0.25 rad of apocentre. Across the 32,000 distinct initial states, $a_{\rm c}$ spans 0.20649--0.21674 au with a median of 0.21165 au, while $e_{\rm c}$ spans 0.11159--0.16002 with a median of 0.13576 (Fig.~\ref{fig:initial_conditions}). We describe this population as dynamically cold in the transfer sense: its orbits are selected to give low relative speeds during exchange between neighbouring planets, although the comet eccentricities are not close to zero. Initialisation near apocentre samples each orbit before its first planetary encounter. Candidates lying within three mutual Hill radii of a planet in either member of a paired calculation are rejected.

Each of the 16 planetary architectures receives 2000 massless comets at each eccentricity, giving 64,000 end-to-end trajectories. A comet reaches the inner region when its instantaneous stellar distance first crosses inward through the orbit of P2. The calculation continues after this crossing, so reaching the inner region and striking P1 are recorded as separate events.

For $N_{\rm inj}$ injected comets we define
\begin{align}
 P_{\rm reach} &= \frac{N_{\rm reach}}{N_{\rm inj}}, &
 P_{\rm impact} &= \frac{N_{\rm impact}}{N_{\rm inj}}, \\
 Y_{15} &= \frac{N(\vimp<15\,\kms)}{N_{\rm inj}}.
 \label{eq:diagnostics}
\end{align}
The first two quantities measure transport and collision. $Y_{15}$ measures the complete gentle-delivery efficiency, including comets that never reach P1. The $15\,\kms$ threshold summarises the velocity range in which HCN survival changes rapidly for kilometre-scale impactors and has been used in atmospheric-entry and impact-chemistry studies \citep{ToddOberg2020,Anslow2025,McDonald2025}.

The yield can be separated into three physical stages,
\begin{equation}
 Y_{15}=P({\rm reach})\,
 P({\rm impact}\mid{\rm reach})\,
 P(\vimp<15\,\kms\mid{\rm impact}).
 \label{eq:factorisation}
\end{equation}
This form distinguishes supply to the inner system, collision with P1 after arrival, and the velocity quality of that collision.

\subsection{Controlled final-transfer experiments}
\label{sec:controlled_methods}

The end-to-end calculation measures the net outcome of many successive encounters. A second set of integrations isolates the final transfer to P1. Comets are initialised between P1 and P2 with the same dimensionless pericentre, apocentre and phase window in equation~\eqref{eq:comet_window}. Each condition contains eight planetary architectures and 400 comets per architecture.

Two separations, $\Delta=10$ and 30, are integrated at $e_{\rm p,0}=0$, 0.01, 0.03 and 0.05. This grid measures how the impact-speed distribution and its lowest-speed tail respond to eccentricity in compact and wider transfer geometries. Since the circular minimum impact speed differs with spacing, the common low-speed diagnostic for this experiment is
\begin{equation}
 F_{+1}=P\!\left[\vimp<v_{\rm imp,min}(\Delta)+1\,\kms\right],
 \label{eq:fplusone}
\end{equation}
where $v_{\rm imp,min}(\Delta)$ is the analytic circular-orbit minimum set by the transfer geometry and the planet's escape speed. The $1\,\kms$ interval is small compared with the planet's $11.18\,\kms$ escape speed and matches the order of the epicyclic velocity introduced by a few per cent eccentricity at P1. Referencing this narrow interval to the minimum at each spacing isolates the velocity-coherent edge of each distribution.

For $\Delta=10$, the adjacent period ratio is close to 3:2. We therefore repeat the circular and $e_{\rm p,0}=0.05$ calculations at $\Delta=12$, whose period ratio is 1.638. This comparison tests the eccentricity response across adjacent period ratios of 1.507 and 1.638.

\subsection{Numerical integration and encounter recording}
\label{sec:numerics}

The integrations use \textsc{rebound} 4.4.10 \citep{ReinLiu2012}. The hybrid MERCURIUS scheme advances regular motion symplectically and switches close encounters to the adaptive IAS15 integrator \citep{ReinSpiegel2015,Rein2019}. The fixed symplectic step is $P_1/50$, where $P_1$ is the orbital period of P1. End-to-end runs last 3000 P8 periods; controlled final-transfer runs last 50 outer-planet periods. A comet is removed when it collides with the star or a planet, or when its stellar distance exceeds $100a_8$.

The star and planets are assigned the radii in Table~\ref{tab:parameters}; comets are point particles. A planetary collision occurs when the comet--planet separation reaches the physical planetary radius. The relative speed recorded by the collision callback is corrected to exact surface contact by two-body energy conservation. For an Earth-mass, Earth-radius target,
\begin{equation}
 \vimp^2=v_\infty^2+v_{\rm esc}^2, \qquad
 v_{\rm esc}=11.18\,\kms,
 \label{eq:impact_speed}
\end{equation}
where $v_\infty$ is the comet--planet excess speed before close gravitational acceleration and $v_{\rm esc}$ is the planet's surface escape speed. This escape speed is the lower bound approached by a comet with negligible $v_\infty$.

To locate the origin of the velocity shift, a detection surface is placed at the instantaneous Hill-radius scale
\begin{equation}
 R_{\rm H}=a_{\rm p}\left(\frac{m_{\rm p}}{3M_\star}\right)^{1/3}.
 \label{eq:hillradius}
\end{equation}
When a comet crosses inward through this surface, the code stores its relative velocity
\begin{equation}
 \ventry=\left|\boldsymbol v_{\rm c}-\boldsymbol v_{\rm p}\right|,
 \label{eq:ventry}
\end{equation}
its radial and tangential components, and the osculating eccentricity-vector difference $|\boldsymbol e_{\rm c}-\boldsymbol e_{\rm p}|$. Entries separated by an excursion outside the Hill sphere are recorded individually, and the entry that leads directly to a collision is identified. The impact angle $\theta$ is measured from the local horizontal: $0^\circ$ is grazing and $90^\circ$ is vertical.

\begin{table}
\centering
\caption{Physical parameters of the fiducial end-to-end architecture.}
\label{tab:parameters}
\begin{tabular}{lc}
\toprule
Parameter & Value \\
\midrule
Stellar mass, $M_\star$ & $0.1\Msun$ \\
Stellar radius, $R_\star$ & $0.12\mathrm{R}_{\odot}$ \\
Planet mass, $m_{\rm p}$ & $1\Mearth$ \\
Planet radius, $R_{\rm p}$ & $1\Rearth$ \\
Inner semimajor axis, $a_1$ & 0.0355 au \\
Outer semimajor axis, $a_8$ & 0.24037 au \\
Planet separation, $\Delta$ & 10 \\
Initial planet eccentricity, $e_{\rm p,0}$ & 0, 0.05 \\
\bottomrule
\end{tabular}
\end{table}

\subsection{Statistics and HCN survival-equivalent yield}
\label{sec:statistics}

The independent statistical unit is a planetary architecture. Differences and confidence intervals are obtained from 10,000 paired bootstrap resamples of the 16 end-to-end architectures. The controlled experiments use the same paired resampling over eight architectures. Event counts are also reported for the rare P1 impacts.

As a chemical application, each end-to-end impact is mapped to the equilibrium-chemistry fit of \citet{McDonald2025}. Their parametrisation gives the percentage of HCN surviving an idealised impact, $\gamma(\vimp,\theta,D)$, as a function of surface impact speed, angle from the horizontal and comet diameter. We set $D=1$ km and calculate
\begin{equation}
 Y_{\rm HCN}=\frac{1}{N_{\rm inj}}
 \sum_{i=1}^{N_{\rm impact}}
 \frac{\gamma(v_{{\rm imp},i},\theta_i,D)}{100}.
 \label{eq:hcn_yield}
\end{equation}
Thus, one impact with 20 per cent HCN survival contributes 0.2 surviving-comet equivalents before division by the number injected. The quantity combines impact frequency, velocity and angle in one downstream measure.

All active planetary systems remain free of collisions and orbit crossing over their respective integrations. The maximum relative energy error in the end-to-end calculation is $1.33\times10^{-7}$; smaller-step and all-IAS15 calculations show no systematic shift in the impact-speed distribution.

\section{Results}
\label{sec:results}

\subsection{Transport through the planet chain}
\label{sec:transport_results}

Table~\ref{tab:fates} shows how the fate of the injected population changes between circular and eccentric chains. In the circular systems, 34.1 per cent of comets strike P8. At $e_{\rm p,0}=0.05$ this fraction falls to 30.4 per cent, an 11 per cent relative reduction. Impacts on P2--P6 rise from 25.2 to 26.5 per cent, and the surviving population at 3000 outer periods rises from 11.3 to 14.1 per cent. Ejection remains uncommon, increasing from 0.18 to 0.25 per cent; no comet strikes the star.

The change at the outer edge permits more trajectories to penetrate the chain. The fraction entering the inner region increases from 3.56 to 3.78 per cent, a 6.3 per cent relative increase. Paired trajectory histories identify the origin of much of this gain: among comets that reach the inner region only in the eccentric calculation, 77 per cent strike P6--P8 in the paired circular calculation. Eccentricity weakens interception during the first stages of transfer and opens additional paths to smaller radii.

\begin{table}
\centering
\caption{Final states after 3000 P8 periods. The two middle columns give percentages of all injected comets; the final column gives the relative change from $e_{\rm p,0}=0$ to 0.05.}
\label{tab:fates}
\begin{tabular}{lrrr}
\toprule
Final state & $e_{\rm p,0}=0$ & $e_{\rm p,0}=0.05$ & Change \\
 & (per cent) & (per cent) & (per cent) \\
\midrule
P8 impact & 34.13 & 30.40 & $-10.9$ \\
P7 impact & 28.78 & 28.37 & $-1.5$ \\
P2--P6 impacts & 25.15 & 26.45 & $+5.2$ \\
P1 impact & 0.409 & 0.400 & $-2.3$ \\
Ejected & 0.181 & 0.250 & $+37.9$ \\
Surviving & 11.35 & 14.13 & $+24.6$ \\
\bottomrule
\end{tabular}
\end{table}

The fraction striking P1 is nearly unchanged: 0.409 per cent in the circular chain and 0.400 per cent at $e_{\rm p,0}=0.05$. The enhanced flux into the inner region therefore does not produce a corresponding increase in the innermost planet's total impact rate.

\subsection{Impact velocities and gentle-delivery yield}
\label{sec:yield_results}

The velocity distribution separates the two cases despite their similar impact rates. Among P1 impacts, the fraction below $15\,\kms$ decreases from 25.2 per cent for circular planets to 18.0 per cent for $e_{\rm p,0}=0.05$. After normalisation by all injected comets, the central value of the gentle-impact yield changes from
\begin{equation}
 Y_{15}=1.031\times10^{-3}
 \quad\hbox{to}\quad
 0.719\times10^{-3},
 \label{eq:y15_result}
\end{equation}
a reduction of 30.3 per cent. The raw samples contain 131 and 128 P1 impacts, of which 33 and 23 meet the velocity threshold.

\begin{figure}
\includegraphics[width=\columnwidth]{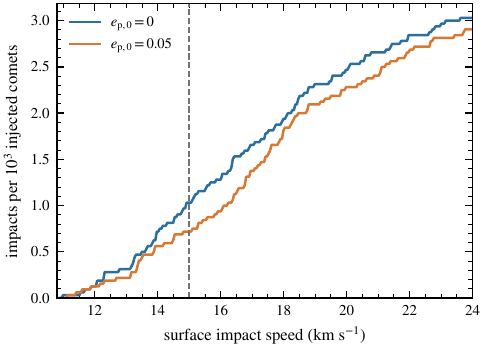}
\caption{Cumulative end-to-end P1 impact yield per $10^3$ comets injected between P7 and P8. The dashed line marks $15\,\kms$. The normalisation includes comets that fail to reach or strike P1, so the curve measures the complete delivery pathway.}
\label{fig:yield}
\alttext{The eccentric-planet curve lies below the circular-planet curve at low and intermediate impact speeds, including at $15\,\kms$.}
\end{figure}

The cumulative yields in Fig.~\ref{fig:yield} diverge at the slow end and remain separated through the central part of the distribution. Architecture-level bootstrap intervals are narrow relative to the line width over much of the plotted range. For the scalar $Y_{15}$ ratio, the paired 95 per cent interval is 0.40--1.11 and 92 per cent of paired bootstrap resamples give a lower yield in the eccentric systems.

Equation~\eqref{eq:factorisation} gives the physical origin of the net change. From $e_{\rm p,0}=0$ to 0.05, $P({\rm reach})$ rises by 6.3 per cent. Among the comets that reach the inner region, the fraction that subsequently strikes P1 decreases by 8.1 per cent. Among P1 impacts, the fraction below $15\,\kms$ decreases by 28.7 per cent. Their product gives
\begin{equation}
 \frac{Y_{15}(0.05)}{Y_{15}(0)}
 =1.063\times0.919\times0.713=0.697.
 \label{eq:ratio_product}
\end{equation}
The increased supply to the inner region is smaller than the loss of low-speed collisions within the impact sample.

\begin{table}
\centering
\caption{End-to-end delivery to P1. Probabilities are normalised by 32,000 injected comets in each condition.}
\label{tab:endtoend}
\begin{tabular}{lcc}
\toprule
Quantity & $e_{\rm p,0}=0$ & $e_{\rm p,0}=0.05$ \\
\midrule
$P_{\rm reach}$ (per cent) & 3.559 & 3.784 \\
$P_{\rm impact}$ (per cent) & 0.409 & 0.400 \\
$P_{15\mid\rm impact}$ (per cent) & 25.2 & 18.0 \\
$Y_{15}$ ($10^{-3}$) & 1.031 & 0.719 \\
$N_{\rm impact}$ & 131 & 128 \\
$N_{15}$ & 33 & 23 \\
\bottomrule
\end{tabular}
\end{table}

\subsection{Velocity at the Hill-sphere boundary}
\label{sec:hill_results}

\begin{figure*}
\includegraphics[width=\textwidth]{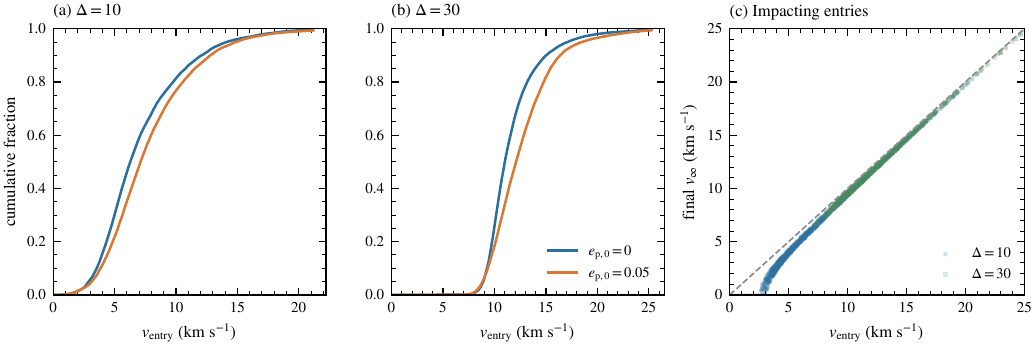}
\caption{Comet--planet relative velocity at entry to P1's Hill sphere in the controlled final-transfer experiment. Panels (a) and (b) show cumulative distributions for $\Delta=10$ and 30. Panel (c) contains entries that lead directly to P1 impact and compares entry speed with the final excess speed derived at the surface. The dashed line marks equality.}
\label{fig:hill}
\alttext{The Hill-entry speed distributions shift to higher velocities at $e_{\rm p,0}=0.05$, while impacting entries retain a nearly one-to-one relation between entry speed and final $v_\infty$.}
\end{figure*}

The controlled final-transfer integrations show where the slow impacts are lost. At $\Delta=10$, increasing the initial planet eccentricity from 0 to 0.05 raises the median P1 Hill-entry speed by $0.82^{+0.36}_{-0.39}\,\kms$. At $\Delta=30$, the shift is $1.09^{+0.19}_{-0.20}\,\kms$ and the interval between the 10th and 90th percentiles broadens by $1.15\,\kms$. The cumulative distributions in Figs~\ref{fig:hill}a and \ref{fig:hill}b show that the deficit is strongest among the slowest entries.

Entries that proceed directly to collision preserve their velocity ordering (Fig.~\ref{fig:hill}c). The excess speed inferred at the surface through equation~\eqref{eq:impact_speed} follows $\ventry$ closely; the small curved offset at low speed reflects the stellar and planetary potential between the Hill boundary and the asymptotic two-body definition of $v_\infty$. The impact-speed shift is therefore established before close gravitational focusing.

The local orbital state supplies the predictive velocity scale. At $e_{\rm p,0}=0.05$, $\ventry$ correlates strongly with $|\boldsymbol e_{\rm c}-\boldsymbol e_{\rm p}|v_{\rm K}$, with rank coefficients of 0.82 at $\Delta=10$ and 0.52 at $\Delta=30$. Correlations with the median relative eccentricity among neighbouring planets are close to zero. The encounter is governed by the eccentricity-vector difference of the comet and target planet at that time, after selection by all preceding planet-to-planet transfers.

\subsection{Eccentricity and planet spacing}
\label{sec:grid_results}

\begin{figure*}
\includegraphics[width=\textwidth]{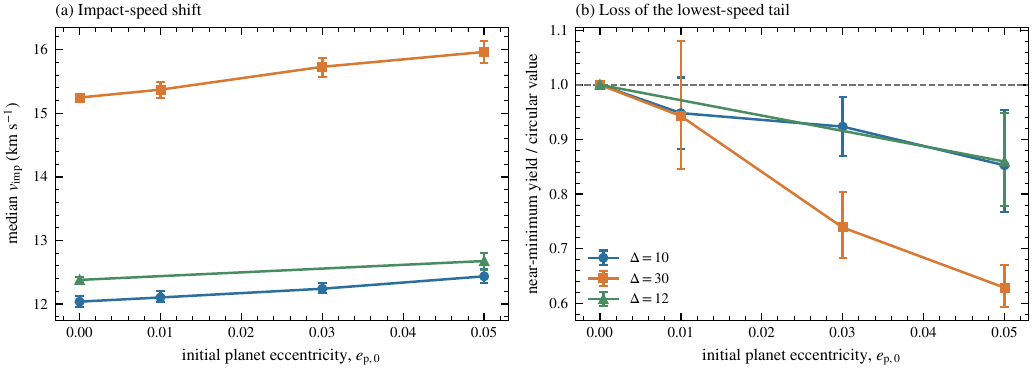}
\caption{Controlled final-transfer experiments. Left: median P1 surface impact speed as a function of initial planet eccentricity. Right: yield within $1\,\kms$ of the circular analytic minimum, normalised to the circular value at each spacing. Points are means over eight planetary architectures; error bars are paired 95 per cent bootstrap intervals. The $\Delta=12$ calculation tests the response away from the near-3:2 period ratio of the $\Delta=10$ chain.}
\label{fig:grid}
\alttext{Increasing planet eccentricity raises the median impact speed and lowers the near-minimum-speed yield at $\Delta=10$, 12 and 30.}
\end{figure*}

The velocity response develops across the observed few-per-cent eccentricity range (Fig.~\ref{fig:grid}). At $e_{\rm p,0}=0.01$, the median impact speed increases by $0.07\,\kms$ for $\Delta=10$ and $0.12\,\kms$ for $\Delta=30$. At 0.03 the shifts are 0.20 and $0.48\,\kms$; at 0.05 they reach 0.40 and $0.72\,\kms$. The wider planet spacing begins with a higher circular encounter speed and shows the larger absolute shift.

For $\Delta=10$, the P1 impact probability stays near 0.41 while the yield within $1\,\kms$ of the circular minimum falls by 15 per cent between $e_{\rm p,0}=0$ and 0.05. At $\Delta=30$, the impact probability falls from 0.262 to 0.230 and the corresponding near-minimum yield falls by 37 per cent. Eccentricity first erodes the most velocity-coherent encounters; wider spacing also reduces the number of impacts.

The $\Delta=12$ systems respond in the same direction. At $e_{\rm p,0}=0.05$, their median Hill-entry speed increases by $0.67^{+0.21}_{-0.19}\,\kms$, median impact speed increases by $0.30^{+0.16}_{-0.18}\,\kms$, and the near-minimum yield falls by 14 per cent. The trend therefore persists after the adjacent period ratio is moved from 1.507 to 1.638, away from the near-3:2 geometry of the $\Delta=10$ chain.

\section{Discussion}
\label{sec:discussion}

\subsection{Why inward transport and gentle delivery diverge}

The simulations separate two consequences of planetary eccentricity. At the outer edge, eccentric planets sample a wider range of radii and velocity directions during each orbit. Comets that would meet P8 in the circular geometry can pass through its encounter zone, increasing the population available to middle planets and the number that reaches the inner region. This redistribution resembles the broader result from planet-chain scattering studies that inner-system flux, planetary accretion and survival are distinct outcomes of the same transport process \citep{Marino2018}.

Near P1, a gentle impact requires close alignment of the comet and planet velocity vectors. The circular compact chain repeatedly selects this geometry because successive transfers require modest changes in orbit. Planetary eccentricity adds radial motion and changes the tangential speed at a given orbital radius. It also changes which comet trajectories survive the preceding planets. The low-speed part of the Hill-entry distribution is depleted, and gravitational focusing then adds the same Earth-like escape-speed scale to both eccentricity conditions. The total P1 impact probability remains stable because the supply of inward-moving comets has increased, while their velocity distribution has become less coherent.

This explains the factorisation in equation~\eqref{eq:factorisation}. Transport quantity rises, collision efficiency after arrival declines slightly, and impact quality shows the largest change. A flux measured at an inner boundary therefore cannot be converted directly into a prebiotic delivery rate without the subsequent encounter and collision velocities.

\subsection{Compact systems with modest eccentricity}

The tested eccentricities occupy the range inferred for many compact multi-planet systems. Transit-timing analyses find typical eccentricities of a few per cent \citep{HaddenLithwick2017}, while transit-duration studies find a broader but still dynamically cold distribution for multi-transiting systems \citep{VanEylen2019}. In this range, eccentricity need not destabilise the chain during the delivery episode. It can still alter the relative eccentricity vectors that control close encounters, consistent with the stability behaviour found for tightly packed eccentric systems \citep{GratiaLissauer2020}.

The effect is amplified by the orbital scale of late M-dwarf temperate zones. At $v_{\rm K}\simeq50\,\kms$, $e=0.03$ corresponds to $1.5\,\kms$ of epicyclic motion. The analytic circular transfer at $\Delta=10$ has $v_\infty\simeq3.3\,\kms$, so an eccentric contribution of this size changes the lowest-speed encounters appreciably. The controlled grid begins to show a clear displacement near $e_{\rm p,0}=0.03$, and the $\Delta=12$ experiment gives the same response at a second period ratio.

Planet count and mutual-Hill spacing remain important predictors of inward scattering \citep{BonsorWyatt2012,Wyatt2017}. The present results add the instantaneous comet--planet eccentricity-vector difference as a local predictor of the final encounter speed. Architectures with the same spacing can deliver different low-speed fractions when their eccentric motions differ.

\subsection{Consequences for prebiotic delivery}

The HCN calculation translates the dynamical shift into a chemical quantity. Equation~\eqref{eq:hcn_yield} weights every P1 impact by the survival fraction predicted from its speed and angle for a 1-km-diameter comet. The circular and eccentric chains have nearly identical numbers of P1 impacts, yet $Y_{\rm HCN}$ decreases from $3.53\times10^{-4}$ to $2.43\times10^{-4}$ surviving-comet equivalents per injected comet. The central reduction is 31 per cent and follows the loss of slow, favourable-angle events.

HCN is relevant to several prebiotic pathways because it can be stored in ferrocyanide salts and later feed the synthesis of nucleobases, amino acids and lipid precursors \citep{Patel2015,McDonald2025}. The result links planetary architecture to the chemical inventory available at a rocky surface: a system that delivers more icy bodies inward can still supply less intact feedstock through impacts. Around low-mass stars, where close-in temperate planets already face high characteristic impact speeds, modest eccentricity further narrows the set of chemically favourable events. Orbital dynamics therefore contributes to the boundary conditions for habitability through the amount and state of material delivered, alongside stellar radiation, atmospheric evolution and surface geochemistry \citep{Rimmer2018}.

\subsection{Limitations and future work}
\label{sec:limitations}

The comet population is introduced on transfer orbits between the two outermost planets. A debris belt, resonances and any planets exterior to the model determine how frequently such orbits are supplied and their distribution in Tisserand parameter, inclination and phase. A self-consistent outer reservoir is the next step for converting the conditional yields measured here into rates for individual systems.

The sensitivity to the source population can be estimated from the relative eccentricity of a comet and planet. To leading order,
\begin{equation}
 v_{\rm rel}\sim v_{\rm K}e_{\rm rel}, \qquad
 e_{\rm rel}^{2}=e_{\rm c}^{2}+e_{\rm p}^{2}
 -2e_{\rm c}e_{\rm p}\cos\Delta\varpi,
 \label{eq:relative_eccentricity}
\end{equation}
where $\Delta\varpi$ is their apsidal separation. If comet pericentres are randomly oriented, $\langle e_{\rm rel}^{2}\rangle\simeq\langle e_{\rm c}^{2}\rangle+e_{\rm p}^{2}$. For a highly excited source with $e_{{\rm c},{\rm rms}}\gg e_{\rm p}$, the fractional change introduced by planet eccentricity is only
\begin{equation}
 \frac{\Delta e_{\rm rel}}{e_{\rm rel}}
 \sim \frac{1}{2}\left(\frac{e_{\rm p}}{e_{{\rm c},{\rm rms}}}\right)^2.
 \label{eq:hot_source_limit}
\end{equation}
Source excitation therefore sets the strength of the eccentricity effect. For the dynamically cold transfer population studied here, the low-speed impact yield falls by 30 per cent. Future end-to-end calculations can establish how the response changes across source distributions produced by outer reservoirs and resonant injection.

The fiducial absolute yield applies to one coplanar architecture with a $0.1\Msun$ star, Earth-mass planets and $\Delta=10$. Further simulations can examine different stellar and planet masses, non-uniform spacings and three-dimensional systems. Observed compact systems provide natural targets for that extension, especially when eccentricity constraints and an outer debris reservoir are both available.

The simulations report finite-time yields at 3000 outer periods. Longer integrations and dedicated rare-event sampling can refine the confidence interval on end-to-end gentle impacts. Coupling these dynamics to atmospheric entry \citep{Anslow2025} and impact chemistry \citep{McDonald2025} across a comet-size distribution will convert survival-equivalent yields into delivered surface masses. These extensions are left for future work.

\section{Conclusions}
\label{sec:conclusions}

We investigated how modest planet eccentricity changes comet delivery through a compact chain extending from a volatile-rich outer region to an Earth-mass planet on a temperate M-dwarf orbit. Paired end-to-end integrations measured the final yield, and controlled final-transfer calculations located the velocity mechanism.

At $e_{\rm p,0}=0.05$, early interception by the outermost planet decreases and the inner-region arrival probability rises by 6.3 per cent. The P1 impact probability changes by less than 3 per cent. The fraction of those impacts below $15\,\kms$ falls from 25.2 to 18.0 per cent, reducing the end-to-end gentle-impact yield by 30.3 per cent in the central estimate.

The slow encounters are removed at the Hill-sphere boundary. Median entry speeds increase by $0.67$--$1.09\,\kms$ across $\Delta=10$, 12 and 30, and the subsequent surface speed follows from this entry condition plus gravitational focusing. The same response at $\Delta=12$ extends the result beyond the near-3:2 geometry of the fiducial chain.

The HCN survival-equivalent yield decreases by 31 per cent even though the number of impacts remains nearly fixed. Planetary eccentricity leaves the inward conveyor active and changes the character of its output: more comets penetrate the chain, while a smaller fraction reaches the innermost planet under conditions favourable to fragile prebiotic material.

\section*{Software}

The simulations and analysis used Python 3.10.12, \textsc{rebound} 4.4.10 \citep{ReinLiu2012}, \textsc{numpy} 2.2.6 \citep{Harris2020}, \textsc{scipy} 1.15.3 \citep{Virtanen2020}, \textsc{pandas} 2.3.3 \citep{McKinney2010} and \textsc{matplotlib} 3.10.7 \citep{Hunter2007}.

\section*{Acknowledgements}

The author thanks Richard J. Anslow for helpful discussions and for sharing simulation material that informed the development of this study. Generative-AI tools were used to assist with code implementation and language editing. The scientific analysis, interpretation, and conclusions were independently validated by the author, who takes full responsibility for the content of this work.

\section*{Data availability}

The simulation code, configuration files, aggregated event tables and figure-generation scripts underlying this article are available from Zenodo at \href{https://doi.org/10.5281/zenodo.23073742}{doi:10.5281/zenodo.23073742}.

\balance
\bibliographystyle{mnras}
\bibliography{references}

\label{lastpage}
\end{document}